\documentclass[preprint,11pt]{elsarticle}

\usepackage{amsfonts}
\usepackage{latexsym}
\usepackage{amsmath}
\usepackage{amssymb}
\usepackage{amscd}
\usepackage{epsfig}
\usepackage{graphicx, subfigure}

\begin{document}

\begin{frontmatter}

\title{A new benchmark set for Traveling salesman problem and Hamiltonian cycle problem}
\address[adrs1]{Flinders Mathematical Sciences Laboratory, College of Science and Engineering, Flinders University}

\author[adrs1]{Pouya Baniasadi\corref{corr}}
\ead{pouya.baniasadi@flinders.edu.au}
\cortext[corr]{Corresponding author}
\author[adrs1]{Vladimir Ejov}
\ead{vladimir.ejov@flinders.edu.au}

\author[adrs1]{Michael Haythorpe}
\ead{michael.haythorpe@flinders.edu.au}

\author[adrs1]{Serguei Rossomakhine}
\ead{serguei.rossomakhine@flinders.edu.au}

\begin{abstract}We present a benchmark set for Traveling salesman problem (TSP) with characteristics that are different from the existing benchmark sets. In particular, we focus on small instances which prove to be challenging for one or more state-of-the-art TSP algorithms. These instances are based on difficult instances of Hamiltonian cycle problem (HCP). This includes instances from literature, specially modified randomly generated instances, and instances arising from the conversion of other difficult problems to HCP. We demonstrate that such benchmark instances are helpful in understanding the weaknesses and strengths of algorithms. In particular, we conduct a benchmarking exercise for this new benchmark set totalling over five years of CPU time, comparing the TSP algorithms Concorde, Chained Lin-Kernighan, and LKH. We also include the HCP heuristic SLH in the benchmarking exercise. A discussion about the benefits of specifically considering outlying instances, and in particular instances which are unusually difficult relative to size, is also included.
\end{abstract}


\end{frontmatter}

\section{Introduction and Background}

When analysing the performance of algorithms or developing a new algorithm, it is essential to have good benchmark instances. Benchmark instances are useful for two primary reasons. First, they allow for comparisons with competing algorithms. Second, they stress algorithms in various ways, which helps to identify weaknesses or even potential bugs in the implementation.

Given their importance, it is natural to ask what constitutes a good benchmark set. It seems sensible that a good benchmark set should contain enough variety to indicate the expected performance of the algorithm in general, that is, on instances not contained in the benchmark set. Therefore, a good benchmark set would provide insight on the following:

\begin{itemize}
\item {\bf Generic performance:} What performance can be expected from the algorithm on a generic example.
\item {\bf Weaknesses of an algorithm:} The situations in which the performance of an algorithm deteriorates.
\item {\bf Comparative advantages:} When comparing two algorithms, the situations in which the performance of the two algorithms diverges.
\end{itemize}

One common approach for generating benchmark sets is to randomly generate generic examples through sampling the instance space. Well known examples of such a benchmark set include SATLIB \cite{SATLIB} for boolean satisfiability, and TSPLIB \cite{TSPLIB} for TSP and its variations. This approach is good at identifying the fastest algorithms (including their implementations) on generic examples. However, it has been recognised that this often provides little insight into the weaknesses of an algorithm, or the comparative advantages of competing algorithms \cite{AllWrong}.

In recent times, there has been growing interest in creating benchmark instances by sampling particular sections of the instance space where the performance of different algorithms diverges. This approach first attempts to characterize the instance space by features of the instances (ideally features which can be linked to difficulty), and then identifies which algorithms perform best in different sections of the instance space. If a meaningful set of features is chosen to characterize the instance space, this approach should permit the identification of comparative advantages in competing algorithms. Examples of these studies that rely on machine-learning and statistics include Leyton-Brown et al. \cite{WMD} for winner determination problem, Smith-Miles et al. \cite{kate-tspDifficulty} for Traveling salesman problem, Cho et al. \cite{knapsack1} and Hall and Posner \cite{knapsack2} for knapsack problems, and Smith-Miles et al. \cite{kate-scheduling} for job shop scheduling problem.

The approach which we advocate here is to construct inherently difficult examples with minimal size. That is, small examples that are difficult for many algorithmic approaches. In a sense, this approach focuses on finding or constructing {\bf challenging outlier instances} with the highest level of difficulty for the most competent approaches. This approach is best at exposing the situations in which an algorithm performs poorly. Indeed, the authors have found these insights extremely helpful for identifying how to improve our own algorithms in developmental stage. Furthermore, studying challenging outlier examples could provide clues on important features that induce the difficulty of these examples.

Of course, researchers have long attempted to construct such challenging outlier instances for various algorithms. Perhaps the most famous example is the seminal paper by Selman et al \cite{selman} in which they describe the ideal ratio of clauses to variables in difficult randomly generated instances of boolean satisfiability. However, for many difficult and widely studied problems, similar results are rare or even non-existent. In particular, literature does not currently contain benchmark sets for the Traveling salesman problem (TSP) where small challenging outlier examples are explicitly collected and systematically studied as a benchmark set.

Hence, in this manuscript, we consider TSP, along with the closely related Hamiltonian cycle problem (HCP). Despite the latter also being a widely studied problem, there are practically no recognised benchmark sets for HCP, and the few benchmark instances which do exist are trivial to solve for modern HCP algorithms and hence provide no real insight about their relative performance. We will consider a particular type of benchmark instance, which can be viewed as either an instance of HCP or TSP, and which has a very different character to instances in the existing benchmark sets for TSP. We will show that instances of this type stress the most powerful TSP algorithms in ways that the existing benchmark sets do not, and in doing so expose areas of potential improvement in those algorithms.

Our benchmark set is available on our website, http://fhcp.edu.au/tsphcp \cite{benchmarkingSite}.

\subsection{Traveling Salesman Problem}

The Traveling Salesman Problem (TSP) is defined by the following statement: \emph{Given a set of cities and the cost of travel between each pair of cities, what is the cheapest way of visiting each city exactly once and returning to the origin city?} TSP has attracted the interest of many researchers due to its fundamental theoretical importance, as well as its wide scope of applications, ranging from electronics \cite{electronics-application} to genetics \cite{genetics-application}.

Despite its simple definition, TSP is extremely difficult to solve in general. Indeed, TSP is known to be an NP-hard problem, and even determining if a tour exists of distance no greater than a given upper limit is an NP-complete problem \cite{korte}. However, fast heuristic algorithms that do not guarantee optimal solutions can, nonetheless, obtain optimal or near-optimal solutions for many instances of TSP.

The effectiveness of a TSP heuristic on a given instance relies on many factors such as the size and structure of the instance. Some heuristics may be quite effective at solving a certain class of problems, but perform poorly on a different class of problems. Hence, to fully evaluate the effectiveness of a TSP heuristic, it should be tested on a wide range of instances incorporating various different characteristics. For this purpose, many benchmark sets have been developed. Perhaps the most famous benchmark instance is the 49 city problem (corresponding to capital cities of mainland USA states), which was first solved to provable optimality by Dantzig et al. \cite{Dantzig} in the 1950s. More recently, benchmarking sets such as TSPLIB \cite{TSPLIB}, the World TSP challenge \cite{WorldChallenge}, and the DIMACS Challenge set \cite{DIMACSChallenge} have been developed and are used by developers of TSP algorithms to evaluate their performance. TSPLIB is a set of randomly generated instances of TSP (and its variations) containing up to 85,900 cities. The World TSP challenge contains a number of 2D-TSP instances\footnote{A 2D-TSP instance is a TSP where the citites are embedded on a plane and the distances to be computed as Euclidean distances.} based on satellite data of cities in various countries throughout the world. The DIMACS TSP challenge, first introduced in 2000, includes a set of TSP instances of size up to 10 million cities, consisting primarily of 2D-TSP instances, as well as seven randomly generated instances. When it was first released, DIMACS invited developers of state of the art algorithms to submit their best solutions to these problems. For many of these instances, it is still unknown whether the best known tours are optimal.

All three of these benchmark sets are similar in character, containing either randomly generated instances, or 2D-TSP instances. The randomly generated instances tend to be difficult only because of large size. The 2D-TSP instances are sometimes difficult even at moderately small size; when this is the case, it is usually because the cities are clustered. Indeed, it seems that truly difficult instances require the presence of particular structures.

In this paper, we will consider TSP instances which have entirely different character to those in the recognised benchmark sets. Our approach involves considering the Hamiltonian cycle problem (HCP), which has the following definition: {\em Given a graph $G$, determine whether there exists a simple cycle that includes all vertices in the graph.} Such a simple cycle that includes all vertices in the graph is called a {\em Hamiltonian cycle}, and any graph which contains at least one Hamiltonian cycle is called {\em Hamiltonian}. HCP is known to be NP-complete \cite{karp}, and therefore it encompasses some of the significant difficulty of TSP. There are many ways to convert an instance of HCP to an instance of TSP. However, a simple conversion from HCP to TSP is as follows: set the distance between any two cities $i$ and $j$ equal to 0 if the edge $(i,j)$ exists in $G$, and equal to 1 otherwise. Then, any tour with total distance equal to zero will be optimal, and will correspond to a Hamiltonian cycle in $G$. Therefore we can think of HCP as the binary form of TSP.

Since HCP is equivalent to binary TSP, we can use it to induce graph structure that is very difficult (often impossible) to construct in 2D-TSP instances, and extremely unlikely to occur in randomly generated instances. This in turn provides an opportunity to create new benchmark instances that will stress TSP solvers in ways other benchmark sets will not. There are two additional benefits to generating benchmark problems in this way; first, if the underlying instance of HCP is known to be Hamiltonian, then we know in advance the length of the optimal tour, so we can easily confirm whether a heuristic has been successful in finding it. And second, producing a benchmark set for TSP in this way has the added benefit of also providing a benchmark set for HCP. As mentioned previously, there are very few recognised benchmark sets for HCP, and those few which exist (most famously, the nine HCP instances contained in TSPLIB \cite{TSPLIB}) contain very few instances, of which all are trivial to solve for modern HCP algorithms.

In this paper we compile a list of good candidate HCP instances to be used as TSP benchmark instances. In Section \ref{sec-liter} we report on the best HCP instances from literature for this purpose. In Section \ref{sec-rand} we consider randomly generated HCP instances, and propose a method for iteratively modifying random instances to obtain new instance that may challenge TSP solvers. In Section \ref{sec-npc}, we consider the possibility of converting challenging instances of other NP-complete problems into instances of HCP. Finally, in Section \ref{sec-results} we conclude with an analysis of the results of such a benchmarking exercise on some of the best TSP heuristics, as well as a HCP heuristic, and provide our final thoughts in Section \ref{sec-conclusions}.

\subsection{Properties that contribute to the difficulty of instances}

As mentioned above, the performance of TSP heuristics on instances depends on a number of factors. When comparing two heuristic methods for TSP, one may outperform the other on one class of instances while falling behind on a different class. To analyse the strengths and weaknesses of an algorithm, we require a wide range of benchmarking instances with different characteristics. Obviously, one such factor is size; larger instances are generally more difficult to solve. The difficulty is exacerbated by the additional memory requirements when solving on a standard computer. Most existing benchmark sets focus on the size of the problem as the primary measure of difficulty. For this reason, we have put our focus on other factors that contribute to the difficulty of the problem. Since our instances are all based on an underlying instance of HCP (that is, a graph), we loosely refer to these factors as {\em graph structure}.

We claim that the performance of solvers on certain instances will vary given the presence or absence of certain graph structures. For example, some TSP algorithms can take advantage of symmetry when it is present. Among other graph features that may affect performance of an algorithm are density of edges, repeated structures, and prevalence of Hamiltonian cycles. At this stage, the exact effect of certain graph structures on the performance of TSP algorithms is not well understood, and heuristics can exhibit vastly different performance on two graphs with seemingly similar structure.

As indicated already, in the present work we focus on considering difficult instances with minimal size. We believe that such an approach is potentially more rewarding in terms of identifying weaknesses and strengths of an algorithm in a broader context. In most cases, the instances considered are members of an infinite family, and so larger instances can be constructed if so desired.

\subsection{Algorithms to be tested}

In order to determine which instances provided the greatest challenge for algorithms, we conducted a benchmarking exercise. In that exercise, we used the following four algorithms, including three TSP algorithms, and one specialised HCP heuristic.

\begin{itemize}
\item {\bf Concorde} (v 3.12.19) \cite{ConcordeBook} is arguably the most well-known exact TSP solver. It uses clever linear programming techniques to obtain and verify an optimal solution. Typically, the verification process takes much longer than obtaining the solution.
\item {\bf Chained Lin-Kernighan} algorithm (CLK) (v 3.12.19)\cite{ConcordeBook} is a recent implementation of the famous Lin-Kernighan heuristic algorithm \cite{linkern}, and is included in the Concorde package.
\item {\bf Helgaun's implementation of Lin-Kernighan (LKH)} (v 2.0.7)  \cite{Helsgaun} is a recent implementation of the famous Lin-Kernighan heuristic known for its success of having discovered the best known solutions for many benchmark examples including the World TSP challenge. It holds a number of records for TSP.
\item {\bf Snakes and Ladders Heuristic (SLH)} (v 2.49) \cite{slh} is an efficient heuristic algorithm designed specifically for solving HCP instances.
\end{itemize}

Of course, for many of the instances we tested, we found that all four algorithms were able to solve them trivially. Hence, in the sections that follow, we only include instances (or families of instances) that stressed at least one of the algorithms. In keeping with our desire to consider small instances, all instances considered contain fewer than 10,000 cities, other than a sole exception coming in Section \ref{sec-rand}.

We performed all our tests on a single core of an AMD Opteron(tm) Processor 6282 SE. We have limited our usage of virtual memory to 4 GBs.

\section{HCP instances in literature}\label{sec-liter}

In this section, we summarise some of the most difficult HCP instances that are known in literature, and attempt to solve them using the four test algorithms. It should be noted that these instances were not initially designed to be difficult for HCP heuristics, however their particular structures or features lend themselves to difficulty. For each instance, we produced 100 random relabellings of the vertices, and ran the four algorithms on each of the 100 relabelled instances. We report on the number of times an optimal tour was obtained, as well as the average time it took for the algorithm to run in the cases where it was successful.

\subsection{Generalized Peterson Graphs}

Description: Generalized Peterson Graphs $GP (p,k)$ are a family of 3-regular graphs which are constructed by combining an inner star polygon and an outer regular polygon. Graph $GP (p,k)$ contains $n = 2p$ nodes and $m = 3p$ edges. A precise description is given in \cite{watkins}. Certain choices of $n$ and $k$ result in graphs with special characteristics, in particular graphs with a high degree of symmetry and known number of Hamiltonian Cycles \cite{schwenk}. In Table \ref{tab:GPtypes} we summarise three classes that we focus on here.

\begin{table}[h!]
\footnotesize
\begin{center}
  \begin{tabular}{ | c | c | c | }
    \hline
    $n$ & $k$ & Number of Hamiltonian Cycles \\ \hline \hline
    $n \equiv 1 ($mod $6)$  & $2$ & $n$ \\ \hline
    $n \equiv 3 ($mod $6)$ & $2$ & $3$ \\ \hline
    $n \equiv 5 ($mod $6)$ & $2$ & $0$ \\
    \hline
  \end{tabular}
  \caption{Three types of Generalized Peterson Graphs and their respective number of Hamiltonian cycles.}
  \label{tab:GPtypes}

\end{center}
\end{table}

Note that for $n \equiv 5$ mod 6 and $k = 2$, the graphs are non-Hamiltonian. Furthermore, it is known that these graphs are also hypohamiltonian \cite{bondy}, and hence the addition of any edge will introduce Hamiltonian cycles. Hence, in these cases, we add one edge to create Hamiltonian graphs for benchmarking purposes. We will refer to these three classes of Generalized Petersen graphs as GPN, GP3 and GP0 respectively.

Benchmarking notes: The classes of Generalized Petersen Graphs considered are highly symmetric, and GPN and GP3 instances contain relatively few Hamiltonian cycles. It is not clear how many Hamiltonian cycles the GP0 instances have once an edge is added, however all of them must include that edge. These characteristics make it very difficult for algorithms to take advantage of structures within the graph. Therefore they constitute very difficult examples for most algorithms even when the size of the graph is small, and it can be seen in Table \ref{tab:GP} that all considered algorithms encountered some difficulty solving them, either in terms of unsuccessful runs, or rapidly increasing solving time. As a result, they constitute extremely useful instances.

\setlength\tabcolsep{4 pt}
\begin{table}[h!]
\footnotesize
\begin{center}  \begin{tabular}{|c | c | c | c | c | c | c | c | c | c |}
    \hline
    \bf{Name} & $n$ & \multicolumn{2}{|c|}{\bf{Concorde}} & \multicolumn{2}{|c|}{\bf{CLK}} & \multicolumn{2}{|c|}{\bf{LKH}} & \multicolumn{2}{|c|}{\bf{SLH}} \\
              &      & Solved & Time & Solved & Time& Solved & Time & Solved & Time\\ \hline  \hline
    GPN\_122          &  122    & 0*  & NA  & 0  & NA & 100  & 0.02  & 100 & 0.05 \\ \hline
    GPN\_244          &  244    & 100  & 0.69  & 100  & 0.02 & 100  & 0.02  & 100 & 0.02 \\ \hline
    GPN\_482          &  482    & 0*  & NA  & 0  & NA & 0  & NA  & 100 & 6.54 \\ \hline
    GPN\_998          &  998    & 0*  & NA  & 0  & NA & 0  & NA  & 100 & 101.96 \\ \hline
    GP3\_126          &  126    & 0*  & NA  & 1  & 0.09 & 78  & 0.06  & 100 & 0.31 \\ \hline
    GP3\_246          &  246    & 0*  & NA  & 0  & NA & 1  & 0.22  & 100 & 4.40 \\ \hline
    GP3\_486          &  486    & 0  & NA  & 0  & NA & 0  & NA  & 100 & 51.50 \\ \hline
    GP3\_1002          & 1002     & 0  & NA  & 0  & NA & 0  & NA  & 100 & 1284.23 \\ \hline
    GP0\_130          &  130    & 100  & 0.24  & 100  & 0.07 & 100  & 0.01  & 100 & 0.00 \\ \hline
    GP0\_250          &  250    & 98***  & 223.11 & 36  & 0.20 & 100  & 0.04  & 100 & 0.09 \\ \hline
    GP0\_490          &  490    & 0*  & NA  & 21  & 0.50 & 99  & 0.17  & 100 & 0.45 \\ \hline
    GP0\_1006         &  1006    & 0*  & NA  & 7  & 1.32 & 93  & 1.07  & 100 & 1.97 \\ \hline
  \end{tabular}
    \caption{Comparative performance of the 4 algorithms on generalized Peterson Graphs. Sign `*' next to a number means the reported failures were inability of the process to find the optimal tour within 24 hours. Sign `***' next to a number means that the reported failure were due to the process reporting a bug. All other reported failures indicate the inability of the process to find the optimal tour at the conclusion of the process.}
  \label{tab:GP}
\end{center}\end{table}

\subsection{Sheehan Graphs}

Description: Sheehan \cite{sheehan} described a family of maximally dense uniquely Hamiltonian graphs; that is, Hamiltonian graphs with only a single Hamiltonian cycle, that contain as large a ratio of edges to vertices as possible.

Benchmarking notes: The high density of edges and the fact that the graphs contain only one Hamiltonian cycle make Sheehan graphs especially interesting instances for heuristics that struggle with high density of edges. However, these graphs contain a structural weakness that can be exploited; one at a time, edges can be identified as being impossible to include in a Hamiltonian cycle and removed, until only the Hamiltonian cycle remains. As can be seen in Table \ref{tab:sheehan}, Concorde and LKH, which both make use of linear programming techniques, performed well on this set, while CLK and SLH did not.

\begin{table}[h!]
\footnotesize
\begin{center}  \begin{tabular}{|c | c | c | c | c | c | c | c | c | c |}
    \hline
    \bf{Name} & $n$ & \multicolumn{2}{|c|}{\bf{Concorde}} & \multicolumn{2}{|c|}{\bf{CLK}} & \multicolumn{2}{|c|}{\bf{LKH}} & \multicolumn{2}{|c|}{\bf{SLH}} \\
              &      & Solved & Time & Solved & Time& Solved & Time & Solved & Time\\ \hline  \hline
    SH\_64          &  64    & 100  & 0.86  & 0  & NA & 100  & 0.00  & 100 & 0.12 \\ \hline
    SH\_125          &  250    & 100  & 3.56  & 0  & NA & 100  & 0.02  & 100 & 16.75 \\ \hline
    SH\_250          &  500    & 100  & 47.92  & 0  & NA & 82  & 0.10*  & 70** & 152.23 \\ \hline
    SH\_500          &  1000    & 100  & 46.65  & 0  & NA & 59  & 0.91*  & 24** & 1321.21 \\ \hline
  \end{tabular}
  \caption{Comparative performance of the 4 algorithms on Sheehan Graphs. Sign `*' next to a number means the reported failures were inability of the process to find the optimal tour within 24 hours. Sign `**' next to a number means that the reported failure were due to the process requiring more than 4 GBs of virtual memory. All other reported failures indicate the inability of the process to find the optimal tour at the conclusion of the process.}
  \label{tab:sheehan}
\end{center}\end{table}

\subsection{Modified Flower Snarks}

Description: A Snark is a connected, bridgeless 3-regular graph with chromatic index equal to 4. Most standard definitions also require that the graph have minimum girth 5 \cite{readwilson}. All Snarks are non-Hamiltonian, and many are hypohamiltonian. One infinite family of hypohamiltonian Snarks is the Flower Snarks discovered by Isaacs \cite{flower}. We consider a modification of Flower snarks obtained by adding one edge to the graph to make it Hamiltonian. Empirically, it appears that instances constructed in this way contain exponentially many Hamiltonian cycles. Despite this, the instances prove difficult for some TSP heuristics as the size increases.

Benchmarking notes: Most of the tested algorithms were capable of solving the modified Flower Snarks, but Concorde and CLK were starting to struggle once the number of vertices reached 1004. Conversely, SLH and LKH both found these instances easy to solve. The reuslts are summarised in Table \ref{tab:snark}.

\begin{table}[h!]
\footnotesize
\begin{center}  \begin{tabular}{|c | c | c | c | c | c | c | c | c | c |}
    \hline
    \bf{Name} & $n$ & \multicolumn{2}{|c|}{\bf{Concorde}} & \multicolumn{2}{|c|}{\bf{CLK}} & \multicolumn{2}{|c|}{\bf{LKH}} & \multicolumn{2}{|c|}{\bf{SLH}} \\
              &      & Solved & Time & Solved & Time& Solved & Time & Solved & Time\\ \hline  \hline
    SN\_124          &  124    & 100  & 0.40  & 73  & 0.09 & 100  & 0.01  & 100 & 0.02 \\ \hline
    SN\_252          &  252    & 100  &  0.40 & 100  & 0.22 & 100  & 0.02  & 100 & 0.01 \\ \hline
    SN\_500          &  500    & 100  & 1.22  & 95  & 0.59 & 100  & 0.11  & 100 & 0.13 \\ \hline
    SN\_1004          &  1004    & 98*  & 549.45  & 43  & 1.44 & 100  & 0.48  & 100 & 0.87 \\ \hline
  \end{tabular}
        \caption{Comparative performance of the 4 algorithms on modified Flower Snarks. Sign `*' next to a number means the reported failures were inability of the process to find the optimal tour within 24 hours. All other reported failures indicate the inability of the process to find the optimal tour at the conclusion of the process.}
  \label{tab:snark}
\end{center}\end{table}

\subsection{Fleischner graphs}

Description: Fleischner \cite{fleischner} introduced two families of graphs with minimum degree 4 and a unique Hamiltonian cycle, that are 2-connected and 3-connected respectively. The smallest of these graphs have 170 vertices\footnote{Strictly speaking, the smallest Fleischner graph has 338 vertices, produced by joining two copies of a 169 vertex graph together. However, a 170-vertex graph is produced by replacing the second copy with a single vertex of degree 2, violating the degree 4 requirement but still producing a difficult instance.} and 408 vertices respectively, and larger instances are produced by chaining multiple copies of certain graphs together in a prescribed fashion. The larger instances have $169k$ vertices and $85+323k$ vertices for $k = 2, 3, \hdots$, respectively. The minimum degree being 4 causes difficulty for some heuristics, as it significantly hampers the use of propagation techniques for branch and bound type approaches.

Benchmark notes: As can be seen in Table \ref{tab:Fleischner}, the Fleischner graphs posed an enormous challenge for all tested heuristics, with practically all attempts resulting in failure. These instances demonstrate that minimum degree paired with a low number of Hamiltonian cycles should be considered a measure of difficulty. Indeed, this provides additional motivation to address the, still open, question posed by Fleischner \cite{fleischner} on whether any uniquely Hamiltonian graphs with minimum degree 5 or higher exist.

\begin{table}[h!]
\footnotesize
\begin{center}  \begin{tabular}{|c | c | c | c | c | c | c | c | c | c |}
    \hline
    \bf{Name} & $n$ & \multicolumn{2}{|c|}{\bf{Concorde}} & \multicolumn{2}{|c|}{\bf{CLK}} & \multicolumn{2}{|c|}{\bf{LKH}} & \multicolumn{2}{|c|}{\bf{SLH}} \\
              &      & Solved & Time & Solved & Time& Solved & Time & Solved & Time\\ \hline  \hline
    FLS\_170          &  170    & 0*  & NA  & 0  & NA & 0  & NA  & 1** & 154.18 \\ \hline
    FLS\_338          &  338    & 0*  & NA  & 0  & NA & 0  & NA  & 0** & NA \\ \hline
    FLS\_507           &  507    & 0*  & NA  & 0  & NA & 0  & NA  & 0** & NA \\ \hline
    FLS\_676          &  676    & 0*  & NA  & 0  & NA & 0  & NA  & 0** & NA \\ \hline
    FLS\_845           &  845    & 0*  & NA  & 0  & NA & 0  & NA  & 0** & NA \\ \hline
    FLS\_1014          &  1014    & 0*  & NA  & 0  & NA & 0  & NA  & 0** & NA \\ \hline
    FLS3\_408          &  408    & 0*  & NA  & 0  & NA & 0  & NA  & 0** & NA \\ \hline
    FLS3\_731          &  731    & 0*  & NA  & 0  & NA & 0  & NA  & 0** & NA \\ \hline
    FLS3\_1054          &  1054    & 0*  & NA  & 0  & NA & 0  & NA  & 0** & NA \\ \hline
  \end{tabular}
        \caption{Comparative performance of the 4 algorithms on Fleischner graphs. Sign `*' next to a number means the reported failures were inability of the process to find the optimal tour within 24 hours. Sign `**' next to a number means that the reported failure were due to the process requiring more than 4 GBs of virtual memory. All other reported failures indicate the inability of the process to find the optimal tour at the conclusion of the process.}
  \label{tab:Fleischner}
\end{center}\end{table}

\section{Randomly generated HCP instances}\label{sec-rand}

It is widely known that randomly generated graphs do not typically provide difficult instances of HCP, and indeed, can usually be solved in almost linear time \cite{frieze}. However, for completeness, we consider randomly generated instances here, as well as discussing a procedure for modifying randomly generated graphs to produce more difficult instances.

\subsection{Randomly generated HCP instances}

Description: There are various procedures for generating random graphs with any number of properties. Here, we use the standard method to produce random regular graphs described by Wormald \cite{wormald}, and in particular, we generate 3-regular graphs. These were chosen as they are very sparse, which makes it less likely that there will be many optimal tours, but are still difficult. Indeed, HCP restricted to 3-regular graphs is known to be NP-complete \cite{hcp23hcp,tarjan}. It could be argued that performing tests on sets of random 3-regular graphs provides a good indicator of how an algorithm might be expected to perform on a ``typical" 3-regular graph. However, in practice real-world problems display different properties rarely captured by random graphs. As long as a few conditions in terms of edge density and connectivity are met, random graphs are almost always Hamiltonian. This includes random regular graphs, which are known to almost always be Hamiltonian \cite{robinson} and typically contain many Hamiltonian cycles. Hence, we expect good heuristics to perform well on these graphs, with difficulty only arising as a result of size.

Benchmarking notes: In Table \ref{tab:rand}, the reported time is the average time over 2000 samples for each algorithm and each size. As can be seen, all algorithms performed well on these instances. For this reason, we have not included these instances in our benchmark repository \cite{benchmarkingSite}.

\begin{table}[h!]
\footnotesize
\begin{center}  \begin{tabular}{|c | c | c | c | c | c | c | c | c | c |}
    \hline
    \bf{Size} & \bf{Sample} & \multicolumn{2}{|c|}{\bf{Concorde}} & \multicolumn{2}{|c|}{\bf{CLK}} & \multicolumn{2}{|c|}{\bf{LKH}} & \multicolumn{2}{|c|}{\bf{SLH}} \\
              &      & Solved & Time & Solved & Time& Solved & Time & Solved & Time\\ \hline  \hline
    250          &  2000    & 100  & 1.46  & 100  & 0.78 & 100  & 0.03  & 100 & 0.01 \\ \hline
    500          &  2000    & 100  & 3.82  & 100  & 2.11 & 100  & 0.13  & 100 & 0.04 \\ \hline
    1000          & 2000    & 100  & 9.69  & 100  & 5.26 & 100  & 0.56  & 100 & 0.16 \\ \hline
    2000          & 2000    & 100  & 14.04  & 100  & 12.73 & 100 & 2.42  & 100 & 0.67 \\ \hline
    4000         &  2000    & 100  & 23.71  & 100  & 32.58 & 100  & 10.51  & 100 & 3.13 \\ \hline
    8000          & 2000    & 100  & 48.34  & 100  & 79.75 & 100  & 46.67  & 100 & 14.80 \\ \hline
    16000          &  2000    & 100  & 210.03  & 100  & 234.53 & 99.95  & 209.18  & 100 & 77.29 \\ \hline
  \end{tabular}
      \caption{Comparative performance of the 4 algorithms on randomly generated HCP instances. All reported failures indicate the inability of the process to find the optimal tour at the conclusion of the process.}
  \label{tab:rand}
\end{center}\end{table}

\subsection{Creating difficult instances from random graphs}

Description: As seen above, randomly generated Hamiltonian graphs are typically unchallenging for competent algorithms. However, it is possible to intelligently modify these graphs to produce more difficult instances of arbitrary order. In particular, it is possible to create instances with very few Hamiltonian cycles (possibly only one) with an iterative method starting from a randomly generated Hamiltonian graph. This is achieved by removing carefully selected edges iteratively, where in each iteration the graph remains Hamiltonian but some Hamiltonian cycles are eliminated. The algorithm below describes our method to modify randomly generated Hamiltonian graphs. We will demonstrate that some algorithms struggle to solve many of the resulting instances.

\emph{Algorithm for producing the modified random graph}\\

\begin{itemize}\item[Step 1:] Set $MaximumCount$ to some appropriately large number, ie 100. Set $Count = 0$.
\item[Step 2:] Create random Hamiltonian graph $G$ with a known Hamiltonian Cycle $HC_i$.
\item[Step 3:] Solve the graph $G$ using the chosen solver and find a Hamiltonian cycle $HC_r$. If no Hamiltonian cycle can be found, or if $HC_i$ is equivalent to $HC_r$, go to 3.1; otherwise, go to Step 4.
\begin{itemize}\item[3.1.] Randomly relabel the vertices of $G$, keeping track of how this alters $HC_i$
\item[3.2.] Set $Count = Count + 1$;
\item[3.3.] If $Count > MaximumCount$, then STOP and output $G$. Otherwise go 3.4.
\item[3.4.] Solve the graph again and obtain a new Hamiltonian Cycle $HC_r$.
\item[3.5.] If $HC_i = HC_r$ go back to 3.1., if not go to Step 4.\end{itemize}
\item[Step 4:] If $HC_i$ is not equivalent to $HC_r$, set $Count = 0$ and go to 4.1.
\begin{itemize}\item[4.1.] Find an edge $e$ of $HC_r$ that is not contained in $HC_i$.
\item[4.2.] Remove edge $e$ from $G$.
\item[4.3.] Go back to Step 2.\end{itemize}\end{itemize}

Note that the Hamiltonian Cycle $HC_i$ will always be present in the graph $G$ at each stage. Therefore the graph remains Hamiltonian. Hence, the algorithm will stop once the chosen solver iterates $MaximumCount$ times without managing to find any Hamiltonian cycle other than the known one; or possibly, without finding any Hamiltonian cycle at all. Obviously, instances for which the known Hamiltonian cycle is found every time are trivial, while instances where it often occurs that no Hamiltonian cycle is found are more difficult.

A key point of the above algorithm is that the tours eliminated at each step are influenced by the algorithm being used.  Therefore, if that algorithm is biased towards certain types of tours, those will be the first eliminated, and the remaining tours may, in some sense, be more ``difficult" for that algorithm to discover. In our testing, we found it was almost always the case that a modified random graph which was very difficult for the algorithm used to modify it, would be trivial to solve for all other algorithms. Nonetheless, the above algorithm provides a procedure for generating instances specifically constructed to stress the particular algorithm being tested. Indeed, a measure of robustness for an algorithm is the degree to which it is impervious to this kind of algorithmic attack.

Benchmarking notes: These modified random graphs can be as large as desired, and typically contain only one or very few Hamiltonian Cycles. They are usually very sparse and they contain many nodes of degree two. Concorde, which takes advantage of structure, found the graphs trivial to solve, and in fact performed better on average on the final graphs since they were sparse. SLH never reported any failures, however it often performed significantly slower as the algorithm progressed; this is explained by understanding that SLH works in stages, and only moves to the next stage once the previous has failed. SLH typically solves random graphs in stage 1, but after the graphs were modified it sometimes needed to move to a later stage in order to solve. LKH and CLK often reported failures, in some cases failing to find a tour in all final 100 iterations. Obviously, when producing a benchmark set by this approach, the trivial instances should be discarded and only the difficult instances retained.

In Tables \ref{tab:LKHmod}--\ref{tab:SLHmod} below, we consider each heuristic individually. For various different graph sizes, we produced 2000 graphs using the above algorithm. For each of the 2000 graphs, we set $MaximumCount = 100$, and recorded the number of failures; that is, how often in the final 100 iterations that no Hamiltonian cycle was found. Then, for each size, we report on the average number of failures, the highest number of failures over all graphs of that size, and the percentage of graphs of that size that had no failures in the final 100 iterations.

\begin{table}[h!]
\footnotesize
\begin{center}  \begin{tabular}{|c | c | c | c | c | c |}
    \hline
        \multicolumn{6}{|c|}{\bf LKH} \\ \hline
    \bf{Size} & \bf{Sample} & \bf{Average degree} &\bf{Average Fail} & \bf{Highest Fail} & \bf{Full success} \\ \hline
    250          &  2000    & 3.26 & 0 & 0 & 100\\ \hline
    500          &  2000    &  3.30 & 0.71 & 43 & 76.2\\ \hline
    1000          & 2000    & 3.32  & 11.85 & 88 & 44.51\\ \hline
    2000          & 2000    & 3.36  & 35.60 & 97 & 35.40\\ \hline
    4000         &  2000    & 3.33  & 59.07 & 100& 28.27\\ \hline
  \end{tabular}
        \caption{Performance of the LKH on modified randomly generated HCP instances. All reported failures indicate the inability of the process to find the optimal tour at the conclusion of the process.}
  \label{tab:LKHmod}
\end{center}\end{table}

\begin{table}[h!]
\footnotesize
\begin{center}  \begin{tabular}{|c | c | c | c | c | c |}
    \hline
            \multicolumn{6}{|c|}{\bf CLK} \\ \hline
    \bf{Size} & \bf{Sample} & \bf{Average degree} &\bf{Average Fail} & \bf{Highest Fail} & \bf{Full success} \\ \hline
    250          &  2000    & 3.31 & 28.26 & 96 & 2.71\\ \hline
    500          &  2000    & 3.31 & 65.45 & 99 & 0\\ \hline
    1000          & 2000    & 3.34  & 93.54 & 100 & 0\\ \hline
    2000          & 2000    & 3.35  & 99.96 & 100 & 0\\ \hline
    4000         &  2000    & 3.39  & 100 & 100 & 0\\ \hline
  \end{tabular}
          \caption{Performance of the CLK on modified randomly generated HCP instances. All reported failures indicate the inability of the process to find the optimal tour at the conclusion of the process.}
  \label{tab:CLKmod}
\end{center}\end{table}

\begin{table}[h!]
\footnotesize
\begin{center}  \begin{tabular}{|c | c | c | c | c | c |}
    \hline
            \multicolumn{6}{|c|}{\bf Concorde} \\ \hline
    \bf{Size} & \bf{Sample} & \bf{Average degree} &\bf{Average Fail} & \bf{Highest Fail} & \bf{Full success} \\ \hline
    250          &  2000    & 3.29 & 0 & 0 & 100 \\ \hline
    500          &  2000    &  3.33 & 0 & 0& 100 \\ \hline
    1000          & 2000    &  3.31 & 0 & 0& 100 \\ \hline
    2000          & 2000    & 3.32  & 0 & 0& 100 \\ \hline
    4000         &  2000    &  3.32 & 0 & 0& 100 \\ \hline
  \end{tabular}
  \caption{Performance of the Concorde on modified randomly generated HCP instances.}
  \label{tab:Concordemod}
\end{center}\end{table}

\begin{table}[h!]
\footnotesize
\begin{center}  \begin{tabular}{|c | c | c | c | c | c |}
    \hline
            \multicolumn{6}{|c|}{\bf SLH} \\ \hline
    \bf{Size} & \bf{Sample} & \bf{Average degree} &\bf{Average Fail} & \bf{Highest Fail} & \bf{Full success} \\ \hline
    250          &  2000    & 3.33 & 0 & 0 & 100 \\ \hline
    500          &  2000    & 3.28  & 0 & 0& 100 \\ \hline
    1000          & 2000    & 3.31  & 0 & 0& 100 \\ \hline
    2000          & 2000    & 3.31  & 0 & 0& 100 \\ \hline
    4000         &  2000    & 3.33  & 0 & 0& 100 \\ \hline
  \end{tabular}
  \caption{Performance of the SLH on modified randomly generated HCP instances.}
  \label{tab:SLHmod}
\end{center}\end{table}

\section{HCP instances corresponding to other NP-complete Problems}\label{sec-npc}

As mentioned earlier, HCP is NP-complete, and is hence at least as difficult as any member of the vast set of NP problems. Hence, any instance of another NP problem can, in theory, be converted to an equivalent instance of HCP. It stands to reason that difficult instances of other problems, upon conversion, will result in difficult instances of HCP. It is common for such conversions to result in dramatic growth in size, even though that growth is bounded polynomially. However, in recent times there has been interest in looking specifically for conversions that only result in linear growth in size (e.g. see Creignou \cite{creignou}, Dewdney \cite{dewdney} and Filar et al \cite{karpred}). Since our intention in this paper is to not rely on problem size, all of the conversions we report on here result in only linear growth in the problem size. Although the growth is linear, in some cases there is a significant constant coefficient, so producing difficult instances of small size is non-trivial. Still, this approach often leads to difficult instances of moderate size.

\subsection{Converting computationally difficult problems to HCP}

We consider four NP-complete problems here, described below. Code to perform each of the conversions is available at \cite{fhcpconversion}.

{\bf Chromatic Number Problem (COL)}: The Chromatic Number problem requests the minimum number of colours required to colour the vertices of a given graph, so that no edge in the graph has endpoints with the same colour. The decision variant of the problem asks whether such a vertex colouring is possible for a given number of colours $k$. For a fixed number of colours $k$, the conversion to HCP results in only linear growth. In all tested examples, $k$ was set to 3.

{\bf Generalized Instant Insanity (II)}: Suppose you have $k$ cubes, where each face of each cube has a given colour from a set of $k$ colours. The Generalized Instant Insanity problem asks if it possible to stack these cubes in a column, oriented such that all $k$ colours can be seen along each of the four long faces. An instance with $k = 4$ was originally marketed as \lq\lq Instant Insanity" by Parker Brothers, and the generalized version with arbitrary $k$ was subsequently shown to be NP-complete \cite{robertson}.

{\bf $n$-Queens Problem (QN)}: The $n$-Queens problem asks if, given an $n \times n$ chessboard, it is possible to place $n$ queens in such a way that none of the queens can take any of the other queens using their standard diagonal movements. The $n$-Queens problem is not NP-complete and it is known that the solution is possible for any integer $n \geq 4$. However, the problem is in NP and can hence be converted to HCP.

{\bf Set splitting problem (SSP)}: The Set splitting problem asks, for a given finite universe set $U$, and a family $F$ of subsets of $U$, if there exists a partition of $U$ into two disjoint non-empty subsets $V$ and $W$ such that each entry of $F$ contains at least one element from both $V$ and $W$.

For each of the above problems, we converted four instances into instances of HCP, ensuring we chose instances small enough that the converted graphs had fewer than 10,000 vertices. We then ran each of these instances 100 times for the four test heuristics. As can be seen in Table \ref{tab:Convertion}, CLK and LKH tended to find these instances difficult. Concorde and SLH were able to solve them to a point, but slowed down significantly as the number of vertices increased, with Concorde solving significantly faster than SLH in all tested cases.

\begin{table}[h!]
\footnotesize
\begin{center}  \begin{tabular}{|c | c | c | c | c | c | c | c | c | c |}
    \hline
    \bf{Name} & $n$ & \multicolumn{2}{|c|}{\bf{Concorde}} & \multicolumn{2}{|c|}{\bf{CLK}} & \multicolumn{2}{|c|}{\bf{LKH}} & \multicolumn{2}{|c|}{\bf{SLH}} \\
              &      & Solved & Time & Solved & Time& Solved & Time & Solved & Time\\ \hline  \hline
    COL\_1000          &  1000    & 100  & 5.49  & 0  & NA & 70  & 2.06  & 100 & 34.40 \\ \hline
    COL\_1950          &  1950    & 100  & 10.30  & 0  & NA & 54  & 9.32  & 100 & 311.55 \\ \hline
    COL\_4110          &  4110    & 100  & 42.81  & 0  & NA & 11  & 59.60  & 100 & 4258.60 \\ \hline
    COL\_7998          &  7998    & 100  & 252.79  & 0  & NA & 0  & NA  & 100 & 39071.52 \\ \hline
    II\_1002          &  1002    & 100  & 83.53  & 0  & NA & 0  & NA  & 100 & 43.35 \\ \hline
    II\_1992          &  1992    & 100  & 1573.68  & 0  & NA & 0  & NA  & 100 & 1630.97 \\ \hline
    II\_3972          &  3972    & 0*  & NA  & 0  & NA & 0  & NA  & 100 & 56890.16 \\ \hline
    II\_7932          &  7932    & 0*  & NA  & 0  & NA & 0  & NA  & 0** & NA \\ \hline
    QN\_1044          &  1044    & 100  & 19.82  & 0  & NA & 0  & NA  & 100 & 70.04 \\ \hline
    QN\_1968          &  1968    & 100  & 136.78  & 0  & NA & 2  & 9.80  & 100 & 1612.05 \\ \hline
    QN\_3894          &  3894    & 100  & 2039.75  & 0  & NA & 0  & NA  & 0** & NA \\ \hline
    QN\_8544          &  8544    & 0*  & NA  & 0  & NA & 0  & NA  & 0** & NA \\ \hline
    SSP\_1011          &  1011    & 100  & 30.10  & 0  & NA & 26  & 2.40  & 100 & 63.47 \\ \hline
    SSP\_2007         &  2007    & 100  & 154.96  & 0  & NA & 1  & 9.50  & 100 & 407.81 \\ \hline
    SSP\_4050          &  4050    & 100  & 1393.69  & 0  & NA & 0  & NA  & 100 & 3053.45 \\ \hline
    SSP\_8040          &  8040    & 100  & 12413.90  & 0  & NA & 0  & NA  & 0** & NA \\ \hline
  \end{tabular}
    \caption{Comparative performance of the 4 algorithms on converted computationally difficult problems to HCP. Sign `*' next to a number means the reported failures were inability of the process to find the optimal tour within 24 hours. Sign `**' next to a number means that the reported failure were due to the process requiring more than 4 GBs of virtual memory. All other reported failures indicate the inability of the process to find the optimal tour at the conclusion of the process.}
  \label{tab:Convertion}
\end{center}\end{table}

\section{Analysis of results on the tested algorithms}\label{sec-results}

We now include a short analysis of the four tested algorithms, and their performance on the full benchmark set.

Concorde was very effective at finding optimal tours. It became less effective when the structure of the graph contained significant symmetries combined with low prevalence of optimal tours, but overall it was the most robust of the tested algorithms. Concorde is an exact algorithm, in the sense that it will not stop until an optimal tour is found. However, we have observed that sometimes the program halts on very difficult instances (for example, some generalized Petersen graphs), therefore possibly exposing some bugs in the implementation.

Chained Lin-Kernighan Algorithm is good at tackling random examples, but struggles when the underlying instances have a low number of Hamiltonian Cycles. It's performance was dominated by that of LKH.

LKH is usually considered the best implementation of the Lin-Kernighan method. It lived up to its reputation by outperforming Chained Lin-Kernighan Algorithm in all of our instances. It was the quickest algorithm among the four algorithms and was usually able to output a solution quickly even when the instances were large. However, it struggled when the underlying instances had a low number of Hamiltonian Cycles. Combining this result with that of CLK seems to indicate that Lin-Kernighan type approaches struggle for these kinds of instances.

SLH was very reliable for finding Hamiltonian Cycles even in many of the structurally difficult graphs, with only the Fleischner graphs providing a consistent challenge. However, it was not as fast as Concorde or LKH, and was limited by the large amount of memory required for bigger instances.

\section{Conclusions}\label{sec-conclusions}

We have demonstrated that studying small but challenging outlier examples can provide an insight into the weaknesses of an algorithm. Such insight could lead to finding an area of improvement for the algorithmic approaches, as well as differentiating between the algorithms.

Our approach has been to find difficult instances of HCP, and assume that these will also provide difficult benchmark instances for TSP, or at least stress TSP algorithms in new ways. This is because the structure of HCP allows us to generate examples that are structurally different from examples in the existing TSP benchmark sets. A future study could look at small but difficult instances of TSP in a more general sense. Furthermore, one could look at different problems and study small but difficult instances in other optimisation settings.

Finally, a future endeavour is to theoretically extract the features of the difficult instances. For example, we have witnessed many instances where high level of symmetry combined with low number of Hamiltonian cycles generate difficult instances for various approaches. However, the type of structural symmetry that contributes to difficulty is not theoretically understood, and other instances with seemingly similar symmetrical properties were trivial to solve. Studying what makes these problems difficult could not only provide us information on how to construct better benchmark examples, it can also shed light on the nature of TSP in general.

\bibliographystyle{plain}

\end{document}